\documentclass[journal]{IEEEtran}
\usepackage{amssymb,graphicx,color}

\newcommand{\beq}{\begin{equation}}
\newcommand{\eeq}{\end{equation}}
\newcommand{\bqa}{\begin{eqnarray}}
\newcommand{\eqa}{\end{eqnarray}}

\newcommand{\ea}{{\it et al.~}}

\newcommand{\smallfrac}[2]{\mbox{$\frac{#1}{#2}$}}

\newcommand{\op}[2]{\ket{#1}\bra{#2}}

\newcommand{\hei}{Heisenberg}

\newcommand{\ro}[1]{\left( {#1} \right)}
\newcommand{\an}[1]{\left\langle{#1}\right\rangle}

\newcommand{\red}{\color{black}}
\newcommand{\blk}{\color{black}}

\newcommand{\HL}{{\hei\ limit}}

\input{Qcircuit}

\begin{document}

%
\title{Adaptive Measurements in the Optical Quantum Information Laboratory}
%
%
%

\author{H. M. Wiseman, D. W. Berry, S. D. Bartlett, B. L. Higgins, and G. J. Pryde
\thanks{H.~M.~Wiseman is the corresponding author. Email: H.Wiseman@griffith.edu.au.  Address: Centre for Quantum Dynamics, Griffith University, Brisbane 4111, Australia.  H.~M.~Wiseman was supported by the ARC Centre for Quantum Computer Technology, and the Centre for Quantum Dynamics at Griffith University.} 
\thanks{ D. W. Berry was supported by the ARC Centre for Quantum Computer Technology, the Department of Physics  at Macquarie University, and the Institute for Quantum Computing.}%
\thanks{S. D. Bartlett was supported by the School of Physics, University of Sydney.}%
\thanks{B. L. Higgins and G. J. Pryde were supported by the Centre for Quantum Dynamics at Griffith University.}}

%
%

\markboth{JOURNAL OF SELECTED TOPICS IN QUANTUM ELECTRONICS: 
Quantum Communications and Information Science}%
{Wiseman \MakeLowercase{\textit{et al.}}: Adaptive Measurements in the Optical Quantum Information Laboratory}
%

\IEEEspecialpapernotice{(Invited Paper)}

\date{today}

\maketitle

\begin{abstract}
Adaptive techniques make practical many quantum measurements that would 
otherwise be beyond current laboratory  capabilities. For example: they allow discrimination 
of nonorthogonal states with a probability of error equal to the Helstrom bound; 
they allow measurement of 
 the phase of a quantum oscillator with accuracy approaching (or in some cases attaining) 
 the \HL; and they allow estimation of phase in interferometry with 
 a variance scaling at the \HL, using only single qubit measurement and 
 control. Each of these examples has close links with quantum information, 
 in particular experimental optical quantum information: the first
 is a basic quantum communication protocol; the second has potential application 
 in linear optical quantum computing; the third uses an adaptive protocol inspired by 
 the quantum phase estimation algorithm. We discuss each of these examples, 
 and their implementation in the laboratory, but concentrate upon the last, 
 which was published most recently [Higgins {\em et al.}, Nature vol. 450, p. 393, 2007].
\end{abstract}


%
\IEEEpeerreviewmaketitle

\section{Introduction}

\IEEEPARstart{M}{easurement} of a quantum system has 
conventionally been defined in terms of an observable, 
represented by an Hermitian operator on the system's 
Hilbert space. However, it is now widely recognized that many 
realistic measurements should not be described in such terms \cite{WisMil09}. 
Rather, the formalism of generalized measurements, described by 
a set of positive maps, is required. Such generalized 
measurements are not only necessary for describing realistic detection; 
they also allow for interesting protocols that conventional (projective)
measurements cannot achieve. A simple example is 
unambiguous (but probabilistic) state-discrimination
for nonorthogonal states \cite{Ivanovic1987,Clark2001}. 

A powerful way to generate interesting generalized measurements
from available detectors in the laboratory is by {\em adaptive measurement 
protocols}. By this we mean the following: An incomplete measurement
is made on the system, and its result used to choose the nature of the 
second measurement made on the system, and so on (until the 
measurement is complete). A complete measurement is one which leaves
the system in a state independent of its initial state, and hence containing
no further information of use \cite{WisMil09}. A measurement may be incomplete by 
being a weak (non-projective) measurement on the system as a whole; or
by being a strong (projective) measurement but only on a subsystem; or in 
other ways. All types allow for adaptive protocols.

Adaptive measurements connect to quantum information, in particular 
experimental quantum optical information, in a number of ways. We will 
review three examples in Sec.~II: distinguishing non-orthogonal states; 
optical phase measurement; and interferometric phase estimation. 
The first is a basic protocol in quantum communication, the second has
potential applications in linear optics quantum computing, and the third 
has been inspired by quantum computing algorithm theory. All three
have been realized in the laboratory in recent years \cite{JM07,Arm02,Higgins07}. 
The last, most recent, of these is analysed in detail from a quantum information
perspective in the remaining Sections of the paper, which cover: quantum limits
to phase estimation (Sec.~III); the Quantum Phase Estimation Algorithm  
(Sec.~IV), and its generalization (Sec.~V); and whether adaptive measurements
are necessary in this context (Sec.~VI).

\section{Applications for Adaptive Measurements} 

\subsection{Distinguishing Non-Orthogonal States} \label{sec:distinguish}

The idea of using adaptive measurements for discriminating between two non-orthorgonal quantum states was introduced as early as 1973 \cite{Dol73}. 
``Dolinar's receiver'' is an optical technique, applicable to 
a traveling mode prepared in one of two possible 
coherent states. The object is to discriminate the preparations with minimum 
probability of error. The minimum possible error probability 
is known as the Helstrom bound \cite{Helstrom}, which in this 
case is $(1/4)e^{-|\Delta \alpha|^2}$, where $\Delta \alpha$ is the difference between
the coherent amplitudes of the two states.  In this case the Helstrom measurement could be realized
 simply by measuring a suitable observable on the harmonic oscillator Hilbert space.
 However, this observable does not correspond to any of the observables usually measured
 in quantum optics, such as a quadrature, or a displaced photon number operator. Indeed,
 the obvious scheme of measuring the quadrature $\hat X_\theta$, with $\theta = \arg(\Delta \alpha)$,
 gives a probability of error of scaling as the square root of the Helstrom bound for large
 $|\Delta\alpha|$ \cite{Dol73}. 
 
 Surprisingly, by  using adaptive detection, one can precisely achieve the Helstrom bound \cite{Dol73}.
 One must measure the leading segment of the 
 pulse, obtain a result, and use that result to alter the measurement on the next segment of the pulse, 
 and so on. \red The Dolinar receiver requires taking the continuous time limit for these segments of pulse, 
 but reasonable results can be obtained as long as the pulse mode has  a 
 duration long compared to the delay in the feedback loop \cite{Ger04}. \blk
 The Dolinar scheme requires measuring a displaced photon number operator (using a weak local oscillator and a photon counter), and altering the  displacement whenever a photodetection occurs.
Very precise control over the applied displacements is required, and very fast electro-optics. 
For these reasons the ``Dolinar receiver'' was not realized experimentally until 2007, by Geremia and co-workers \cite{JM07}. This experiment showed clearly the improvement over the most obvious 
non-adaptive technique for mean photon numbers between $0.1$ and $1$. 

The Dolinar receiver is most naturally and simply described as an adaptive scheme based on 
weak measurements, with the system being one of two coherent states (e.g. $\ket{\pm \alpha}$) of a single-mode harmonic oscillator, of spatial duration $L$. However, because of the unique properties of coherent states, the system can also be thought of as a series of shorter modes, of length $L/M$, each of which is prepared in the same coherent state $\ket{\pm\alpha/\sqrt{M}}$. In the limit $M\to \infty$, each of these short modes corresponds to the system that is measured at a particular time by the detector. Thus, in this guise, the Dolinar receiver appears as an adaptive scheme based on projective measurements of sub-systems, from an ensemble of identically prepared systems. 

One can take this analysis further. In the $M\to\infty$  limit the subsystems can be treated as qubits. The reason is that the mean excitation number for the adaptively-displaced states scales as $M^{-1}$, and so have support on just the first two number states, $\ket{0}$ and $\ket{1}$, which are the eigenstates of the measured quantity (photon number).\footnote{There is a technical issue in that, according to Dolinar's protocol, the displacement diverges at the initial time if the two states are initially equally likely. 
However, as demonstrated in Ref.~\cite{JM07}, a modified protocol with only a moderately large displacement  obtains the great majority of the improvement offered by Dolinar's protocol.}  If the two undisplaced global coherent states are $\ket{\pm\alpha}$, then to leading 
order the corresponding qubit states are $\ket{0}\pm (\alpha/\sqrt{M})\ket{1}$, and displacement in phase space (by a distance of order $M^{-1/2}$) is   equivalent to a qubit rotation (by an angle of order $M^{-1/2}$), and this rotation makes a measurement in the $\{\ket{0},\ket{1}\}$ basis the desired adaptive measurement. Moreover, the Dolinar scheme can then be derived as a special case of the optimal adaptive scheme for discriminating between two 
non-orthogonal qubit states, when one has multiple copies of them. 

This latter problem, with arbitrary pure qubit states, was solved by Ac\`in \ea \cite{Acin2005}, who gave a simple interpretation of the optimal scheme: it corresponds to making the optimal local Helstrom measurement on each copy, taking into account the probability one assigns to each of the two possible preparation procedures, updated according to all the preceding measurements.  Not only is this conceptually simple procedure the optimal adaptive scheme, it also reaches the Helstrom bound for the entire ensemble (as was known already for the special limit of the Dolinar protocol). That is, contrary to what one might have thought, an entangling measurement across all subsystems is not required. Note also that the restriction to qubits in this quantum information setting is no real restriction; if (unlike in the optical case) one assumes one can make arbitrary projective measurements on a single subsystem then one needs only two basis states to describe a system in one of two possible pure states. 

The situation with mixed states  is, however, much more complicated \cite{Hig-discrim1}. 
For qubits (which is now a meaningful restriction), even in the asymptotic limit there is a clear separation in performance between \begin{enumerate}
\item the Helstrom bound (achieved by the optimal joint measurement); 
\item the optimal adaptive scheme (involving dynamic programming);
\item the locally optimal adaptive scheme (i.e. the locally optimal Helstrom measurement); and 
\item the obvious non-adaptive scheme (a majority vote from repeated unbiased measurements).
\end{enumerate}
Moreover, surprisingly, the latter two actually reverse order (i.e. the locally optimal scheme performs worse than the non-adaptive scheme) for a sufficient degree of mixture. 

\red 
Before leaving the topic of adaptive measurements for state discrimination we should mention the 
work of Jacobs \cite{Jac07} on adaptive continuous  measurements on a single qubit 
 prepared in one of two non-orthogonal states $\ket{\phi_\pm}$. 
The continuous measurement limit can be thought of (as in the Dolinar case) as a sequence
of weak measurements of duration $\Delta t$, whose disturbance also scales as $\Delta t$, 
and taking the limit $\Delta t \to dt$. In this case, if one measures for an infinitely long time then one can realize the Helstrom bound for distinguishability simply by making the same weak measurements 
at every step, of the Pauli operator proportional to $\op{\phi_+}{\phi_+} - \op{\phi_-}{\phi_-}$. 
However for some finite times, more information can be obtained by using a locally optimized  
adaptive measurement in which the measured Pauli operator
is continually rotated such that its expected value is zero at all times \cite{Jac07}. At sufficiently long 
 times, this scheme always becomes worse than the simple non-adaptive scheme. It is also worth noting 
 that if one considers the probability of error in one's guess of which preparation was performed, 
rather than the mutual information between one's record and the preparation, then the 
simple non-adaptive scheme always performs better \cite{ComGamWis09}.
\blk

\subsection{Optical Phase Measurement} \label{sec:optical_phase}

Although the Dolinar receiver was the first notable adaptive measurement scheme
to be introduced theoretically, it was not the first to be realized experimentally. 
That honour goes to the adaptive phase measurement algorithm introduced 
by Wiseman and Killip \cite{WisKil97,WisKil98}, and realized by Mabuchi 
and co-workers \cite{Arm02}. This technique is based on homodyne detection, 
rather than photodetection as in the Dolinar receiver. Its aim is not state discrimination 
(although it can be useful for that \cite{WisKil98}), but rather measuring a physical quantity:
the phase of the state. The optimal measurement to do this, a canonical phase measurement, 
would give a measurement result $\phi$ with variance equal to the intrinsic phase variance
of the state, by definition \cite{LeoVacBohPau95}. However, this cannot be realized by 
standard optical measurements. For a state with initially completely unknown phase, 
the best standard technique is heterodyne detection \cite{LeoVacBohPau95}. 
This introduces an excess phase variance scaling as $1/4\bar{n}$, where $\bar{n}$ 
is the mean number of photons in the state \cite{WisKil97}. This scaling is known 
as the {\em standard quantum limit} (SQL). 

If one knew that the phase of the state was approximately $\varphi$, one could 
make a homodyne measurement of the quadrature $\hat X_{\theta}$ with $\theta = \varphi+\pi/2$, 
and this would be almost as good as a canonical phase measurement for many types of 
states.\footnote{The $\pi/2$ phase difference between $\theta$ and $\varphi$ here follows from 
a convention regarding the phase introduced by a beam-splitter, which is generally used in 
papers on this topic, e.g. Refs.~\cite{Wis95c,WisKil97,WisKil98,BerWis01a,Arm02}.} The intuitive idea of an adaptive homodyne measurement is to begin with homodyne 
measurement of a random quadrature, and then to adjust the local oscillator phase $\theta$ 
adaptively over time to $\varphi(t)+\pi/2$. Here $\varphi(t)$ is an estimate of the system 
phase based on the homodyne data so far. Interestingly, one does not want to choose 
$\varphi(t)$ to be the {\em best} estimate of $\phi$ at time $t$ --- that actually gives inferior 
performance for some states \cite{BerWis01a}. At the end of the measurement, one does 
want to chose the final estimate $\phi$ to be the best estimate. For the ``Mark II'' scheme 
of Wiseman and Killip \cite{WisKil97}, the excess phase variance scales as 
$1/8\bar{n}^{3/2}$, far smaller than that of the best non-adaptive measurement. More complicated
adaptive schemes can do even better \cite{BerWis01a}, near the ultimate (Heisenberg) 
limit scaling of $1/\bar{n}^2$. 

The Mark II scheme was experimentally implemented using small, coherently excited, microwave-frequency sidebands of a large coherent beam \cite{Arm02}. In this case the phase to be estimated was actually the phase of the microwave excitation, as carried by the optical frequency sidebands. 
An improvement over the best non-adaptive scheme was seen for mean photon numbers 
$\bar{n}$ between about 10 and 300. Because the experiment was performed 
with coherent states, it is possible to think of it (like the Dolinar case) as a series 
of projective (quadrature) measurements performed on identically prepared 
weak coherent states. However this would not be the case if the experiment were 
performed with a nonclassical state such as a squeezed state. In that case it would 
be necessary to adopt the more natural description, of a 
succession of weak measurement on a single mode.

It is impossible for an adaptive homodyne measurement to attain the accuracy 
of a canonical phase measurement in general \cite{WisKil98}. However, 
there is a known exception: when the state has support on the $\ket{0}$ and $\ket{1}$ photon 
number states. This case was actually solved in the  paper 
which first proposed and analysed adaptive phase measurement \cite{Wis95c}, 
and is the ``Mark I'' scheme of Refs.~\cite{WisKil97,WisKil98}.
It has the interesting consequence that, given a single photon, it is possible to create deterministically 
an arbitrary superposition of the states $\ket{0}$ and $\ket{1}$. This is done by first 
creating the ``mode-entangled'' \cite{WisVac03} single-photon state $\sqrt{\eta}\ket{0,1}+\sqrt{1-\eta}\ket{1,0}$, and then making a canonical phase measurement on one mode, using the 
Mark I scheme. This yields a completely random result $\phi$ (which emphasizes that in this case there is no approximate initial phase that a nonadaptive homodyne scheme could use),
and collapses the second mode into the state $\sqrt{\eta}e^{i\phi}\ket{0}+\sqrt{1-\eta}\ket{1}$.  Since
$\phi$ is known, the phase of this superposition can then be adjusted to the desired phase.

The ability to create arbitrary superpositions of $\ket{0}$ and $\ket{1}$ photon states 
is of course the ability to create arbitrary photonic qubit states using ``single-rail'' logic 
\cite{Kok07}. It turns out that adaptive phase measurements can also be applied 
at other points in linear optics quantum computing to make it possible to use 
single-rail encoding with resources that are far smaller than was previously thought 
possible, and not substantially larger than the resources required
for conventional ``dual-rail'' encoding \cite{RalLunWis05}. 

It is worth noting here 
that even a ``conventional'' linear optics quantum computation can be regarded as 
an enormous adaptive measurement protocol, since it involves making measurements 
on subsystems (single photons) of an entangled state, and manipulating 
the remainder of the photons in a way that depends upon the results of the prior measurements,
before a final measurement that reveals the result of the computation. This is 
true  of both the circuit architecture and the cluster-state architecture for LOQC \cite{Kok07}. 
Indeed the 2007 four-photon cluster-state experiment from the group of Zeilinger \cite{Jennewein07} is perhaps the first adaptive measurement in which the subsystems were entangled prior to the 
measurement.

\subsection{Interferometric Phase Estimation} \label{sec:IPE}
 
While adaptive optical phase measurement is potentially useful for 
quantum computing, the third and final application we  consider 
is linked to quantum computation in the opposite way. That is, it uses
an algorithm from quantum computing theory to inspire new adaptive 
protocols. Specifically, based on the {\em quantum phase estimation algorithm} (QPEA) 
of Cleve \emph{et al.} \cite{Cleve1998,Nielsen00a}, we have devised 
a new family of adaptive protocols for interferometric phase estimation \cite{Higgins07}. 
Moreover, we have implemented one of these algorithms, demonstrating 
Heisenberg-limited scaling for phase estimation for the first time \cite{Higgins07}.

 \begin{figure}
 \centering
\includegraphics[width=0.4\textwidth]{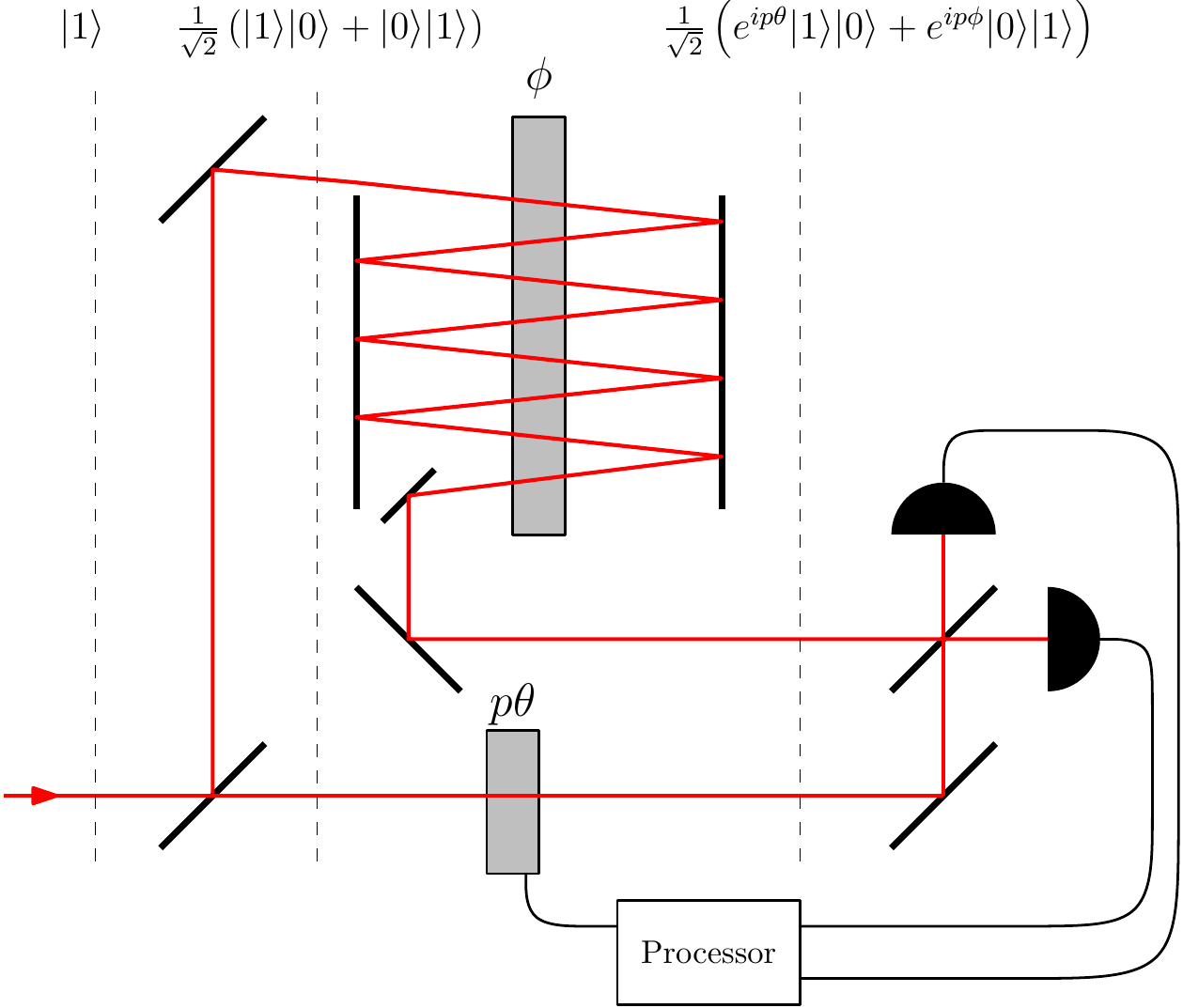} %
\caption{Conceptual diagram of the generalized QPEA implemented using a Mach-Zehnder interferometer, shown with quantum states (expressed in the photon-number basis) at key points. The large phase-shift element is configured to implement an adjustable number $p$ of $\phi$-phase shifts on photons passing through the upper arm (in this example, $p=8$). The small phase-shift element implements an adjustable $p\theta$ phase shift on photons passing through the lower arm. The output of the single photon detectors determines, via the processor, how to adjust $\theta$ prior to the next photon input, and also the final phase estimate $\phi_{\rm est}$.}
\label{fig:multpass-concept}
\end{figure}

The bulk of this paper is dedicated to explaining adaptive (and non-adaptive) 
measurements for interferometric phase estimation from a quantum information  
perspective. In this section we will concentrate on placing it in the context of 
optical interferometry (although the idea could work equally well with particles 
--- such as neutrons \cite{Hra96} --- other than photons). A conceptual experimental
diagram is shown in Fig.~\ref{fig:multpass-concept}. The key differences from 
a standard Mach-Zehnder interferometer are
\begin{enumerate}
\item The number of passes of one arm through the unknown phase shift can be greater than one, and is assumed 
controllable over the course of the experiment.
\item The phase shift in the other arm is assumed controllable over the course of the experiment.
\end{enumerate}
In standard interferometry, $N$ indepedendent photon-detections allow the unknown phase to be estimated with accuracy $\Delta \phi = 1/\sqrt{N}$ (for large $N$), which is known as the 
standard quantum limit (SQL). By contrast, the Heisenberg limit (HL) is quadratically better: 
$\Delta \phi = \pi/N$. 

Those familiar with the QPEA may be surprised to find that applying a generalization of this algorithm
to the Mach-Zehnder interferometer (as will be explored in later sections) yields only 
a quadratic improvement in the accuracy. The QPEA is supposed to give a binary phase readout, 
implying an uncertainty exponentially small in the size of the register. Also, it is at the heart of 
Shor's algorithm \cite{Sho94} which gives an exponential speed-up over classical algorithms for 
factoring. In quantum computing theory, the QPEA is used to estimate the phase of an eigenvalue $e^{i\phi}$ of a (typically multi-qubit) unitary operator $U$ that corresponds to some calculation. If one can do a quantum computation implementing $U$, then one can do (more or less) with the same resource cost a quantum computation implementing $U^p$, for any $p$. Thus in a quantum algorithm context, the number of ``passes'' $p$ is irrelevant.

If we were to follow an analogous method of resource counting in interferometry --- count simply the number of {\em photons} irrespective of the number of times $p$ each photon 
passes through the unknown phase shift --- then 
the algorithm implemented in Ref.~\cite{Higgins07} would have yielded a $\Delta\phi$ 
exponentially small in $N$. This violates the long-established Heisenberg limit scaling \cite{Holevo,Helstrom} with $\Delta \phi$ of order $1/N$. The 
(correct) Heisenberg scaling is obtained by counting not photons, but rather the total number of {\em photon-passes} through the unknown phase shift \cite{GLM06} .  That is, 
one should count each photon involving a $p$-pass interferometer (as shown in Fig.~\ref{fig:multpass-concept}) as using $p$ resources, not 1.

\begin{figure}
 \centering
\includegraphics[width=0.5\textwidth]{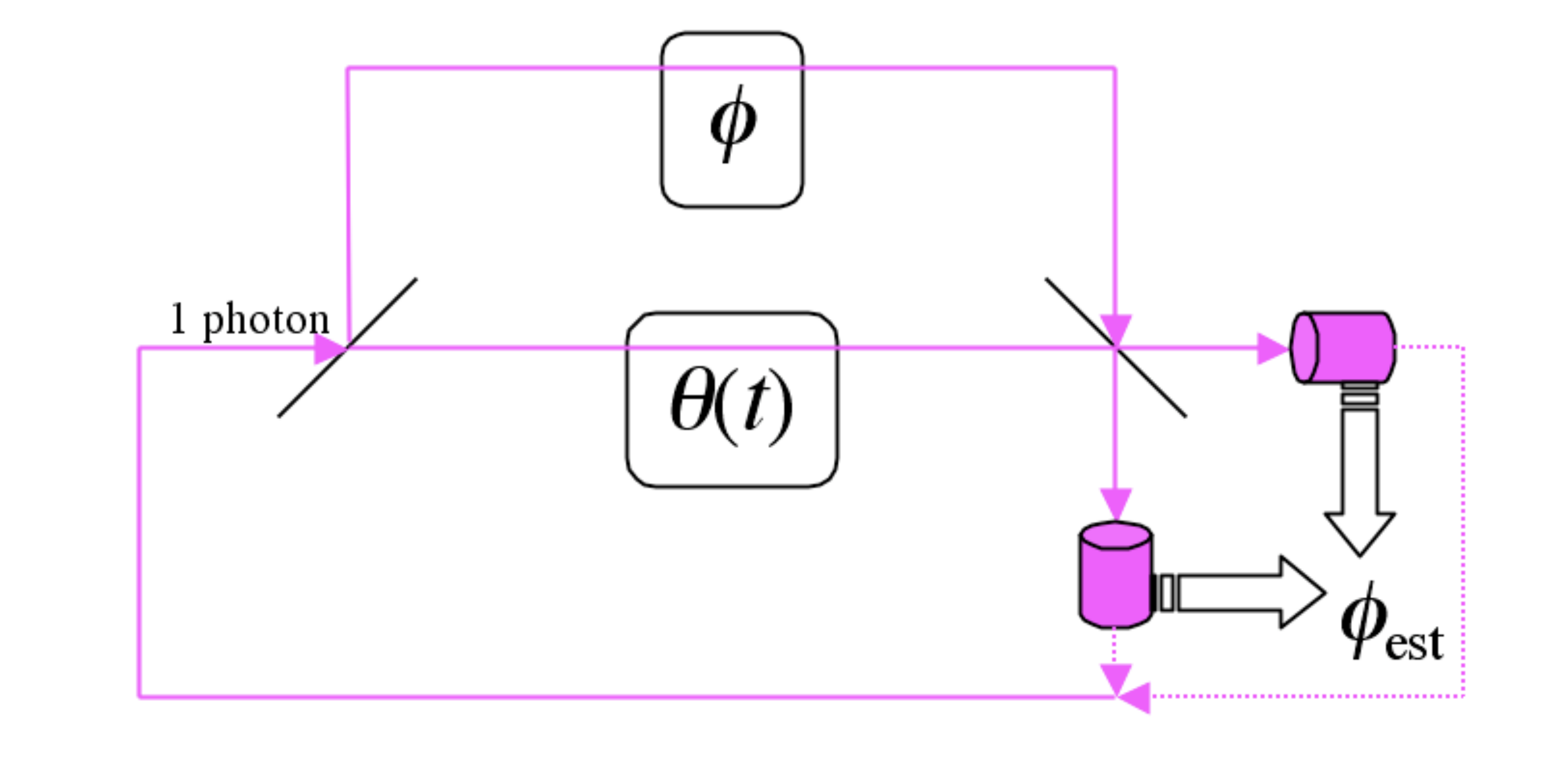} %
\caption{Arbitrarily accurate phase estimation may be done with a single photon if QND measurements are allowed. After detection, the photon (wherever it is found) is redirected back into the interferometer.}
\label{fig:MZI-QND}
\end{figure}

This method of counting is not justified merely by giving the expected Heisenberg-limit. Rather, it has a number of other justifications. First, photon number is not a sensible resource. For example, if one allows for non-demolition photon-number measurements, one could repeat a standard interferometry experiment arbitrarily many times, thus obtaining an arbitrarily small uncertainty, using just a single photon, as shown in Fig.~\ref{fig:MZI-QND}. Other nonlinear optical processes allow for scenarios where it is not even clear how to define the number of photons  \cite{Higgins09}. Second, one practical reason for caring about photon number is if one has a sample that is extremely sensitive to light. In this situation what is relevant is clearly the number of photon-passes through the sample, not anything else. Third, if one counted photons rather than photon-passes, this would ignore the extra time it takes for a photon to make $p$ passes, rather than 1 pass, through the sample. In the asymptotic regime of arbitrarily large $N$ (and hence large $p$), this is necessarily the time that will determine the duration of the experiment. 

It might be thought that the technique of using $p$ passes of a single photon is not really a measurement of $\phi$ at all, but rather a measurement of $p\phi$. From this one would of course expect a sensitivity in $\phi$ that scales as $p$ from a single photon measurement, and so an overall uncertainty in $\phi$  scaling inversely with the total number of photon passes is just as expected. 
This works only if one already knows the phase approximately, and the deviations one is trying to detect are much smaller than $2\pi/p$. This is quite different from the fundamental problem of estimating a completely unknown phase $\phi$, which is the task to which the Heisenberg limit pertains \cite{Holevo,Helstrom}. In the latter case, one cannot simply set $p$ to be as large as it can be, and then measure $p\phi$, because this would yield information only about $\phi$ modulo $2\pi/p$. Rather, even if one sets $p$ to high values at some stage of the experiment, at some other stage(s) it must be set to one in order to pin down a single value for $\phi$ within the range $[0,2\pi)$. That is to say, for the task we are interested in, the measured quantity is the variable $\phi$ modulo $2\pi$, not the variable $p\phi$ modulo $2\pi$. The optimal way to vary $p$ over the course of the experiment is then the crucial issue and the QPEA suggests an answer (or at least a starting point \cite{Higgins07}.) 

In many practical examples, the strategy of fixing $p$ at a value as large as possible works because one is trying to detect small changes to an already known phase.
This is the case in gravitational wave interferometry, and atomic clocks. In the former case, the maximum number of passes is set by the geometry of the experiment and the quality of the mirrors, and this determines the accuracy \cite{LIGWD00}.   In the latter case, the decoherence time of the atoms determines the maximum number of ``passes'' (Rabi cycles) the atoms can undergo, and the standard quantum limit is determined by this, and the number of atoms in the sample \cite{ColdAtomicClocks01}. In both of these cases one is interested in minimizing the error, and hence one would ideally measure $p\phi$ with a large ensemble $M$ of identical systems (photons or atoms). Thus the overall scaling for the uncertainty would be like $M^{-1/2}p^{-1}$. That is, one does not get scaling at the Heisenberg limit if one fixes $p$ and sets $N=Mp$. In practice one cannot even increase $M$ without limit in these experiments. For gravitational wave interferometry, there is an optimal number of photons for which the noises from photon counting and from radiation pressure balance \cite{LIGWD00}. For atomic clocks, increasing the cloud density similarly leads to collisional energy shifts (i.e. phase noise). In summary, in both of these cases the limits are set by practical, rather than fundamental, considerations.

\begin{figure}
 \centering
\includegraphics[width=0.5\textwidth]{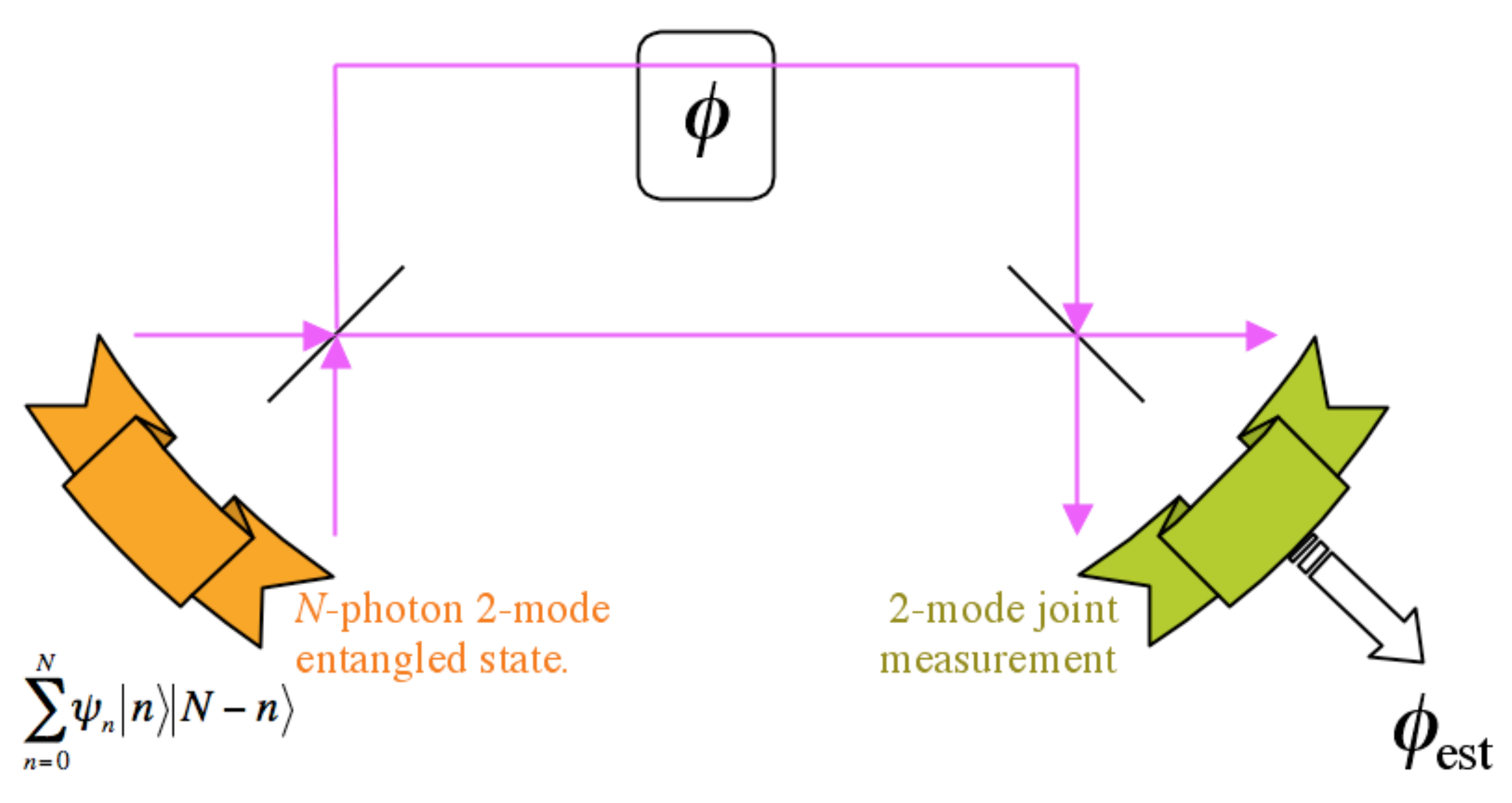} %
\caption{Single-pass Mach-Zehnder interferometer with arbitrary inputs and arbitrary measurements. It is known that this can achieve the Heisenberg-limit $\Delta \phi \sim \pi/N$ for an $N$-photon input state.}
\label{fig:MZI-HL}
\end{figure}

It should be noted that the earlier remarks relating to measuring $\phi$ modulo $2\pi/p$ apply to NOON states with $N=p$ in {\em exactly} the same way as they do to a single photon with $p$-passes. A NOON state is a state of the form $\ket{N,0}+\ket{0,N}$ (expressed using the number basis for the two arms of the interferometer shown in Fig.~\ref{fig:MZI-HL} \cite{Lee2002}). For large $p$, an $N=p$ NOON state has an advantage over a single photon with $p$-passes in that the time it takes to pass through the interferometer does not scale with $p$. It has the obvious disadvantage of being far more difficult to produce --- thus far $N=4$ is the largest NOON state demonstrated (and that only in a post-selected sense) \cite{Nagata07}. It is also far  more difficult to detect --- it requires photon-number-resolving detectors with loss smaller than $O(1/N)$. If it were possible to generate and detect ``high-NOON'' states then the algorithms here could be applied directly to that case as well. The equivalence comes from the fact that all of the phase information in a NOON state after it has passed through the final beam-splitter is contained in the parity of the photon number at one detector (which is why absurdly high efficiencies are required). Note that although the multipass technique can tolerate detector losses of order unity, it is sensitive to absorption within the sample in {\em precisely} the same way as is the NOON state technique.

From the above comparison between NOON states and multiple passes of a single photon, one might  argue that there are really two resources: $N$ (the number of photon passes in both cases); and $T$, the time taken for the experiment. In the limit $N\to\infty$, the duration $T$ for the NOON-state protocol 
increases only logarithmically with $N$ (as explained in the following sections), and for an optimal multiphoton entangled state (see Sec.~\ref{sec:HL}), $T$ need not scale with $N$ at all. 
In the multi-pass case, by contrast, $T$ necessarily scales as $N$, as discussed above. 
In saying that our technique attains the \HL, we mean only 
that it reaches the minimum possible phase variance for a given $N$. That is, to obtain a phase estimate with a binary expansion of $\log_2(N)$ bits, with an error in only the least significant bit, 
we are prepared to accept an experiment of exponentially (in $\log_2 N$) long duration. 
This can be avoided only by moving the exponential cost associated with $N$ from the time 
to the state itself --- the NOON state (and its relatives) contain {\em exponentially many photons} and hence contain an exponentially large amount of energy. In practice, photons are fast and small, so $N$ would have to become very large for either time or energy to become important\footnote{This of course ignores the {\em difficulty} of  converting $N\hbar\omega$ of energy into a NOON state, which is clearly enormous for large $N$, but hard to quantify.}. On the other hand,  the NOON-state protocol also requires an {\em exponentially efficient detector}. As noted above, one requires $1-\eta$ to be exponentially (in $\log_2 N$) small, in contrast to the best current detectors where $1-\eta$ barely qualifies as ``small'' at all. Thus it is hard to see why the resource $T$ should be regarded as more fundamental than these other resources.

\begin{figure}
 \centering
\includegraphics[width=0.5\textwidth]{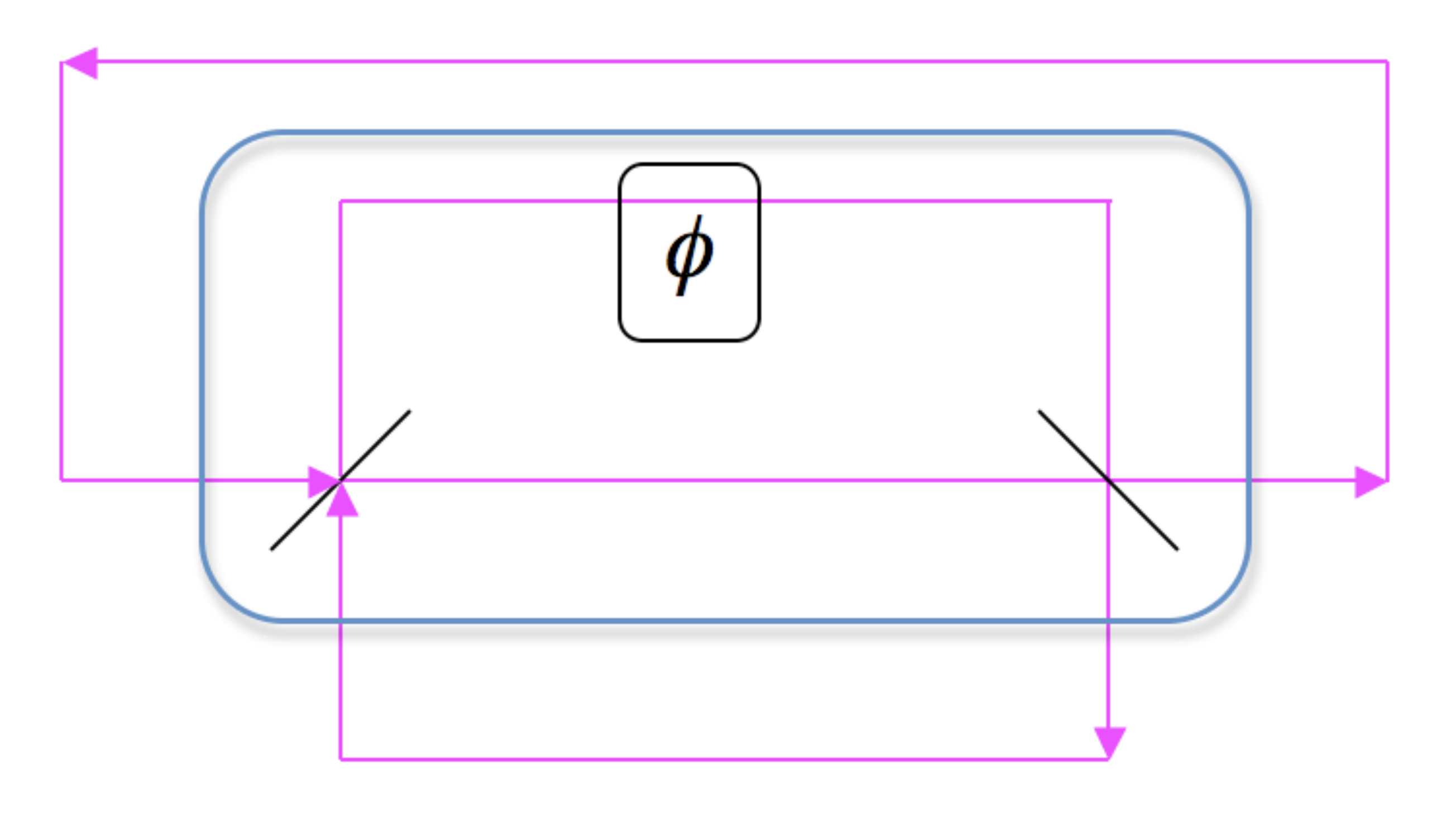} %
\caption{Realising a multipass Mach-Zehnder interferometer from a single-pass one (inside the round-cornered box) by recycling the outputs into the inputs. Note that this is done coherently unlike in Fig.~\ref{fig:MZI-QND}.}
\label{fig:MZI-recycle}
\end{figure}

Finally, it might be thought that a multipass interferometer is somehow changing the rules of the game, in a way that using a NOON state is not. That is, one might argue that the rules allow arbitrary preparation as inputs to the interferometer, and arbitrary processing of the output, as shown in Fig.~\ref{fig:MZI-HL}, but not changing of the beam-paths through the sample.  However it is simple to bypass this objection. If arbitrary preparations and measurements are allowed, then one allowed scheme is not to measure the output modes, but rather simply to redirect them back as the input modes as shown in Fig.~\ref{fig:MZI-recycle}. The input beam-splitter simply undoes the effect of the output beam-splitter, and the second passage of the photon through the unknown phase shift is exactly as in a two-pass interferometer. This can be repeated as many times as desired to give a multipass interferometer. 

\begin{figure}
 \centering
\includegraphics[width=0.5\textwidth]{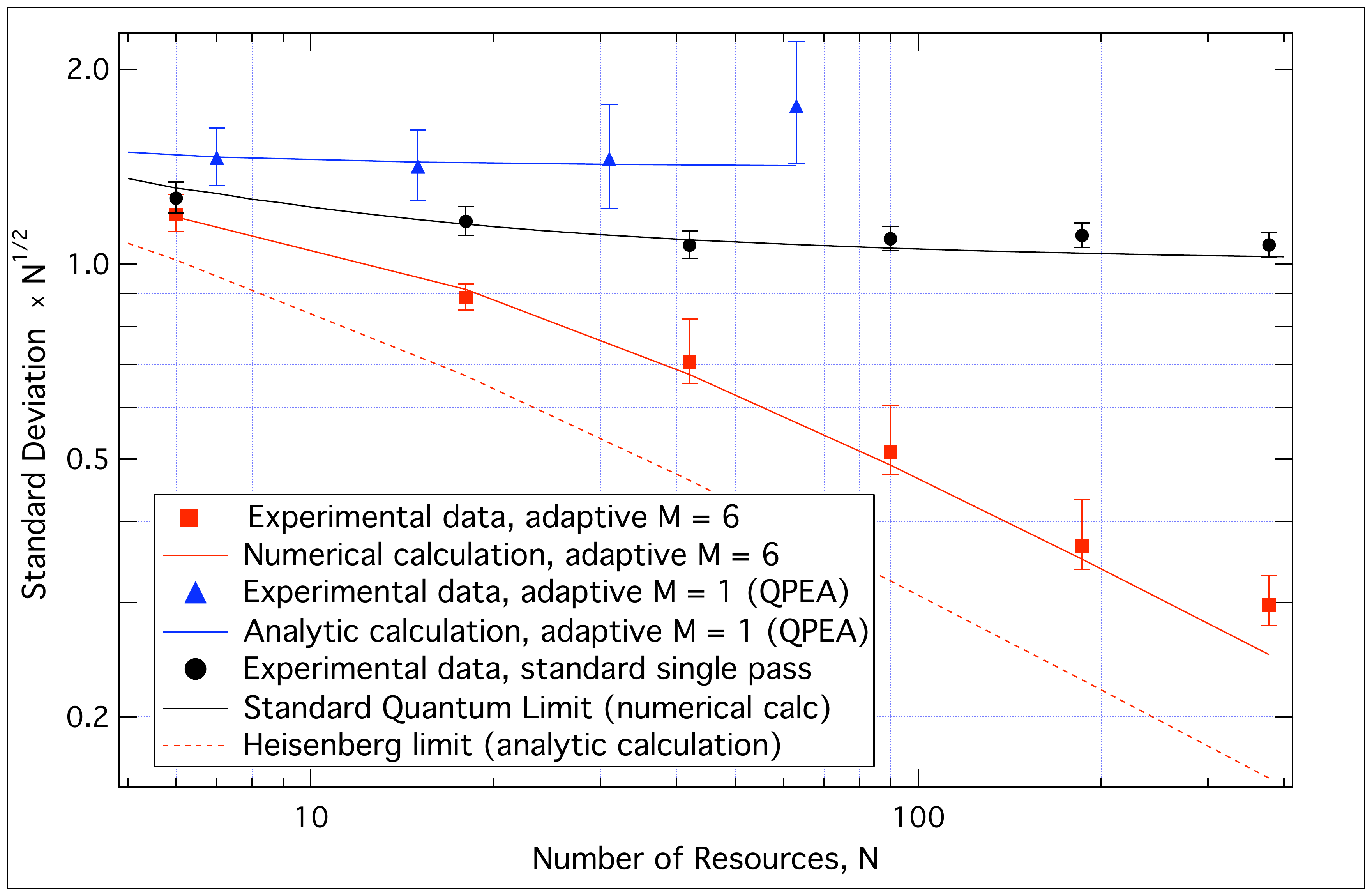} %
\caption{Experimental results  from Ref.~\cite{Higgins07}, of 
standard deviations of phase estimates for varying numbers of resources $N$. We compare theoretical predictions (lines) and measured values (points, each representing 1,000 estimates) for standard phase estimation, the QPEA ($M=1$), and our generalized QPEA ($M=6$) algorithms. Error bars denote 95\% confidence intervals. Our algorithm clearly gives better phase estimates than both the SQL and the QPEA limit.}
\label{fig:results07}
\end{figure}

In Ref.~\cite{Higgins07} we implemented a generalized QPEA using multi-passed single photons with a common-spatial-mode polarization interferometer, which is more stable than a Mach-Zehnder interferometer.  The two arms of the interferometer were the right-circular and left-circular  polarization modes, and the unknown phase $\phi$ was implemented as the orientation angle of a birefringent half-wave plate. The number of passes was varied from 32 to 1, in powers of 2, using electro-mechanical devices. We implemented standard interferometry (verifying the SQL), the QPEA (surprisingly also scaling as the SQL, as will be discussed below), and a generalized QPEA in which $M=6$ photons (rather than $M=1$ as in the QPEA) were sent down consecutively for each pass-configuration. This last technique gave a $\Delta\phi$ scaling the same as the Heisenberg limit, with a multiplicative overhead of only 1.56, as predicted by theory. See Fig.~\ref{fig:results07}.

\section{Quantum Limits to Phase Estimation}

\subsection{Rules and Representations}

In this and following sections we will analyse phase estimation algorithms (adaptive and otherwise) from 
a purely quantum information perspective. That is, we consider general qubits rather 
than photons. In this context, the rules of the game are as follows: 
 \begin{enumerate}
 \item We have a gate that performs the unitary operation $\exp(i\phi \ket{1}\bra{1})$ on a specific sort of qubit.
 \item We have an indefinite supply of these qubits.
 \item The parameter $\phi$ is initially {\em completely unknown}.
 \item We are allowed only $N$ {\em applications of the gate}.
 \item We aim to minimize the variance in our best estimate $\phi_{\rm est}$ of $\phi$.
 \end{enumerate}
 Technically, we use the Holevo variance measure, $V_{H} = \an{\exp[i(\phi-\phi_{\rm est})]}^{-2} - 1$ \cite{Hol84,WisKil97}, as this respects the cyclic nature of phase. The variance is the most robust figure of merit, in that if it scales well then all other measures will also scale well, but not {\em vice versa} \cite{BerWisBre01}.

In quantum information language, a photon entering one port of the interferometer is represented by preparation of the qubit in (say) the logical $\ket{0}$ state. The initial beam splitter acts as a Hadamard transform $H$, yielding $H \ket{0}=\ro{ \ket{0}+\ket{1} }/\sqrt{2} = \ket{+}$. 
The unknown phase shift in the upper arm is represented by the unitary operator $ \exp(i\phi \ket{1}\bra{1})$, while the known phase shift in the lower arm is represented by 
$\exp(i\theta\op{0}{0})$. The final beam-splitter again transforms from the logical ($Z$) basis to the $\ket{\pm}$ basis (the $X$ basis). Thus if the photonic qubit is measured then this amounts to a measurement of $X$ prior to the final beam splitter. Estimating the phase on the basis of passing a single photon once through the interferometer, and measuring it, is therefore described by the circuit 
$$ \Qcircuit @C=+1ex @R=0.3ex @!R @W=12ex {
\lstick{\ket{0}} & \qw & \gate{H} & \qw & \gate{e^{i\theta \ket{0}\bra{0}}} 
& \qw & \gate{e^{i\phi \ket{1}\bra{1}}} &\qw &\measureD{X}
}$$

For compact notation in later circuit diagrams, and also to connect more closely to the QPEA 
of Shor's algorithm, we change to a representation where $\exp(i\phi \ket{1}\bra{1})$ is represented 
by the controlled-unitary gate $\ket{1}\bra{1}\otimes {U}+\op{0}{0}\otimes I$, where $U$ acts on a state $\ket{\phi}$ 
(which could have any Hilbert space dimension) with $ U \ket{\phi} = e^{i\phi}\ket{\phi}$, 
and $I$ is the identity operator on this space.
We also treat the auxiliary phase $\theta$ as a real-number-valued classical register, which controls (indicated by a $\blacklozenge$ symbol) the gate $R(\theta) \equiv \exp(i\theta\op{0}{0})$.  Thus we rewrite the above circuit as 
$$ 
\Qcircuit @C=+1ex @R=0.3ex @!R @W=12ex {
\lstick{\ket{+}}	& \qw&\gate{R(\theta)}	&\ctrl{1}	                &\measureD{X}& \cw \odot                   &          & \\
\lstick{\ket{\phi}}	&\qw    &\qw\cwx			    &\gate{U}	&\qw		        &\qw \cwx			               	& \qw& \\
\lstick{\theta}	   &\cw &\cw\cwx\blacklozenge&\cw 		                &\cw		        &\cgate{D(\delta\theta)}\cwx	&\cw&\rstick{\phi_{\rm est}}\\ 
}
$$
Here if we take $\theta$ to be random
(but known), this ensures an estimate $\phi_{\rm est}$ with an accuracy which is independent of the  true value $\phi$.   Here $\phi_{\rm est}$ is shifted (indicated by the gate $D$) from $\theta$ by an amount 
$\delta\theta$ depending in someway (indicated by $\odot$) on the classical result. The advantage of treating $\theta$ in this way will become apparent. 

\subsection{The Standard Quantum Limit} \label{sec:SQL}

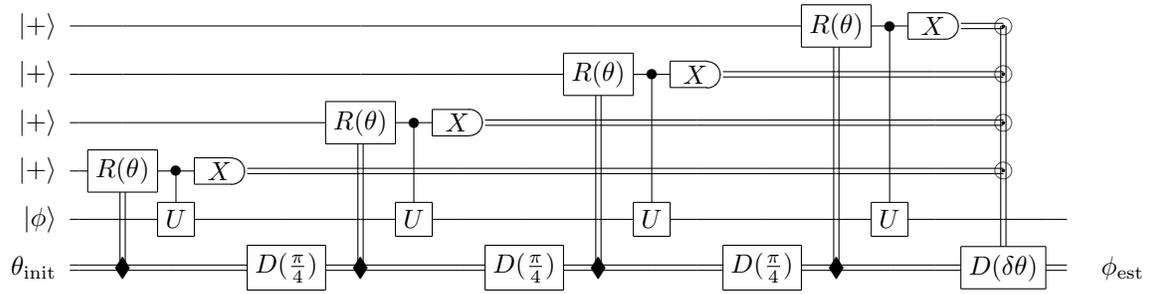
\begin{figure*}
$$
\Qcircuit @C=0ex @R=0.3ex @!R @W=12ex {
\lstick{\ket{+}}	&\push{\rule{1.5ex}{0.4pt}}\qw&\qw			&\qw		&\qw		&\qw					&\qw			&\qw		&\qw		&\qw					&\qw			&\qw		&\qw		&\qw					&\gate{R(\theta)}	&\ctrl{4}	&\measureD{X}	&\odot \cw &&&  \\
\lstick{\ket{+}}	&\qw&\qw			&\qw		&\qw		&\qw					&\qw			&\qw		&\qw		&\qw					&\gate{R(\theta)}	&\ctrl{3}	&\measureD{X}	&\cw				&\cw\cwx		&\cw	 	&\cw		&\odot \cw \cwx &&& \\
\lstick{\ket{+}}	&\qw&\qw			&\qw		&\qw		&\qw					&\gate{R(\theta)}&\ctrl{2}	&\measureD{X}	&\cw				&\cw\cwx		&\cw		&\cw		&\cw				&\cw\cwx		&\cw	  	&\cw		&\cw \cwx\odot &&& \\
\lstick{\ket{+}}	&\qw&\gate{R(\theta)}	&\ctrl{1}	&\measureD{X}	&\cw				&\cw\cwx		&\cw		&\cw		&\cw				&\cw\cwx		&\cw		&\cw		&\cw			&\cw\cwx		&\cw	 	&\cw		&\cw \cwx\odot &&& \\
\lstick{\ket{\phi}}	&\qw&\qw\cwx			&\gate{U}	&\qw		&\qw				&\qw\cwx		&\gate{U}  	&\qw		&\qw			&\qw\cwx		&\gate{U}	&\qw 		&\qw			&\qw\cwx		&\gate{U} 	&\qw		&\qw \cwx				&\push{\rule{0.8em}{0.4pt}\rule{0.8em}{0pt}}\qw& \\
\lstick{\theta_{\rm init}}	&\cw&\cw\cwx\blacklozenge		&\cw 		&\cw		&\cgate{D(\smallfrac{\pi}{4})}	&\cw\cwx\blacklozenge	&\cw		&\cw		&\cgate{D(\smallfrac{\pi}{4})}	&\cw\cwx\blacklozenge	&\cw		&\cw      	&\cgate{D(\smallfrac{\pi}{4})}	&\cw\cwx\blacklozenge	&\cw		&\cw		&\cgate{D(\delta\theta)}\cwx	&\cw&\rstick{\phi_{\rm est}}\\ 
}
$$
\caption{Circuit representation of ``standard'' interferometry with a controllable auxiliary phase. This defines the SQL. Here $N=4$.}
\label{fig:SQLi}
\end{figure*}
 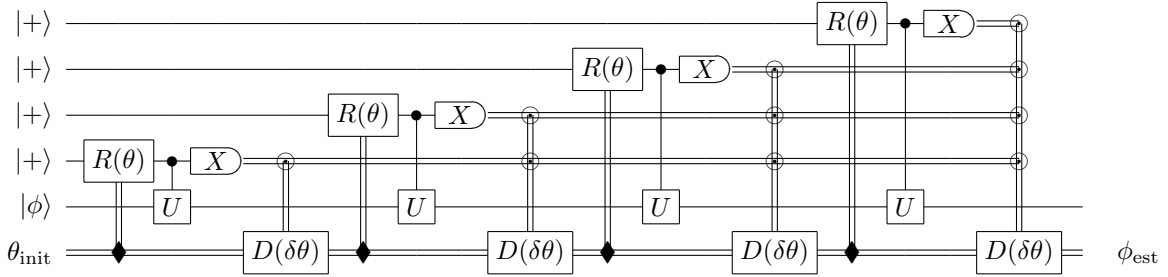
\begin{figure*}
$$
\Qcircuit @C=0ex @R=0.3ex @!R @W=12ex {
\lstick{\ket{+}}	&\push{\rule{1.5ex}{0.4pt}}\qw&\qw			&\qw		&\qw		&\qw					&\qw			&\qw		&\qw		&\qw					&\qw			&\qw		&\qw		&\qw					&\gate{R(\theta)}	&\ctrl{4}	&\measureD{X}	&\odot \cw &&&  \\
\lstick{\ket{+}}	&\qw&\qw			&\qw		&\qw		&\qw					&\qw			&\qw		&\qw		&\qw					&\gate{R(\theta)}	&\ctrl{3}	&\measureD{X}	&{\odot}\cw				&\cw\cwx		&\cw	 	&\cw		&\odot \cw \cwx &&& \\
\lstick{\ket{+}}	&\qw&\qw			&\qw		&\qw		&\qw					&\gate{R(\theta)}&\ctrl{2}	&\measureD{X}	&{\odot}\cw				&\cw\cwx		&\cw		&\cw		&\cw\cwx{\odot}				&\cw\cwx		&\cw	  	&\cw		&\cw \cwx\odot &&& \\
\lstick{\ket{+}}	&\qw&\gate{R(\theta)}	&\ctrl{1}	&\measureD{X}	&{\odot}\cw				&\cw\cwx		&\cw		&\cw		&{\odot}\cw\cwx				&\cw\cwx		&\cw		&\cw		&\cw\cwx{\odot}			&\cw\cwx		&\cw	 	&\cw		&\cw \cwx\odot &&& \\
\lstick{\ket{\phi}}	&\qw&\qw\cwx			&\gate{U}	&\qw		&\qw\cwx				&\qw\cwx		&\gate{U}  	&\qw		&\qw\cwx				&\qw\cwx		&\gate{U}	&\qw 		&\qw\cwx				&\qw\cwx		&\gate{U} 	&\qw		&\qw \cwx				&\push{\rule{0.8em}{0.4pt}\rule{0.8em}{0pt}}\qw& \\
\lstick{\theta_{\rm init}}	&\cw&\cw\cwx\blacklozenge		&\cw 		&\cw		&\cgate{D(\delta\theta)}\cwx	&\cw\cwx{\blacklozenge}	&\cw		&\cw		&\cgate{D(\delta\theta)}\cwx	&\cw\cwx{\blacklozenge}&\cw		&\cw      	&\cgate{D(\delta\theta)}\cwx	&\cw\cwx{\blacklozenge}&\cw		&\cw		&\cgate{D(\delta\theta)}\cwx	&\cw&\rstick{\phi_{\rm est}}\\ 
}
$$
\caption{Circuit representation of ``standard'' interferometry with an adaptively controlled auxilliary phase. The adaptation does not improve the asymptotic accuracy beyond the SQL.}
\label{fig:SQLii}
\end{figure*}

The standard quantum limit pertains when we simply repeat the above single-qubit circuit $N$ times. 
That is, we have ${N}$ qubits, independently prepared, independently measured, and 
with $\exp(i\phi \ket{1}\bra{1})$ applied once on each (${p=1}$).  
To ensure uniform sampling, $\theta_{\rm init}$ is random, and $\theta$ 
is {incremented  by $\pi/N$} between one qubit and the next. The case $N=4$ is shown in Fig.~\ref{fig:SQLi}. This yields the standard quantum limit (SQL) for accuracy, given by 
\beq
{\rm SQL} = V[\phi_{\rm est}] \sim {1/N}  \textrm{ for } N \gg 1.
\eeq
 We can also allow for $\theta$ 
to be {controlled adaptively} between one qubit and the next, as shown in 
Fig.~\ref{fig:SQLii}. The obvious procedure, suggested in Ref.~\cite{BerWis00}, is to choose $\theta$ so as to minimize the {\em expected} variance after the measurement whose result is influenced by this new $\theta$, which entails averaging over the two possible results of the measurement. This local optimization gives a slightly more accurate measurement for small $N$, but  makes no difference asymptotically \cite{BerWisBre01}. A global optimization does even better for small $N$, but again essentially no difference for large $N$ \cite{BerWisBre01}.

\subsection{The \hei\ Limit} \label{sec:HL}

With no restrictions on how the qubits are prepared or measured, an obvious approach is to use $N$ qubits, prepared in a suitable entangled state, and measured by a suitable entangling measurement. In this case the initial and final beam-splitters (Hadamards)  can be absorbed into the state preparation and measurement, and so are irrelevant to the problem. So too is the auxilliary phase. Each qubit controls the phase gate once. For example, with $N=3$ we have: 
$$
\Qcircuit @C=0.5ex @R=0.5ex @!R {
\lstick{\ket{0}} 	&\multigate{2}{\textrm{Ent.~State~Prep.}} &  \qw     			     & \qw     		& \ctrl{3} & \multigate{2}{\textrm{Ent. Meas.}}  		\\
\lstick{\ket{0}}	&\ghost{\textrm{Ent.~State~Prep.}}& \qw     			     &\ctrl{2}	&\qw & \ghost{\textrm{Ent. Meas.}}	   \\
\lstick{\ket{0}}	&\ghost{\textrm{Ent.~State~Prep.}}& \ctrl{1}			     &\qw					&\qw 		        &\ghost{\textrm{Ent. Meas.}}		 \\
\lstick{\ket{\phi}}        &	\qw		  &\gate{U}  &\gate{U}&\gate{U} & \qw  			     &\qw 		&\\
}
$$
Although the phase gates are shown as acting sequentially, they can be imagined to act 
simultaneously. This circuit is equivalent to Fig.~\ref{fig:MZI-HL}, where arbitrary preparation and measurement allows the interferometer to achieve its ultimate performance, the Heisenberg limit, which is given by \cite{BerWis00} 
\beq 
{\rm HL} = V[\phi_{\rm est}] = \tan^2\left(\frac{\pi}{N+2}\right).
\eeq
This is attained by using the canonical phase measurement, described by phase states 
\beq \label{phase_states}
\ket{\Phi} \propto \sum_{n=0}^{N} e^{i\Phi n} \ket{n,N-n}_{\rm S},
\eeq
and the optimal input state \cite{Luis96,BerWis00}, 
\beq
\ket{\psi_{\rm opt}} \propto \sum_{n=0}^N \sin \left[ 
\frac{(n+1)\pi} {N+2}\right] \ket{n,N-n}_{\rm S} \label{psiWB}.
\eeq
Here $\ket{n,N-n}_{\rm S}$ is a symmetrized state in which $n$ qubits are in state $\ket{1}$ and $N-n$ in state $\ket{0}$. (For identical and indistinguishable bosons such as photons, in two modes, this symmetry is enforced by the quantum statistics.) Note that this entangled state differs from the NOON state, which has the form $\ket{N,0}_{\rm S}+\ket{0,N}_{\rm S}$ in this notation. The NOON state gives the maximum  Fisher information (equal to $N$) from a single measurement, but has an appalling variance (the Holevo variance is in fact infinite) because it only detects changes in $\phi$ modulo $2\pi/N$. The optimal state $\ket{\psi_{\rm opt}}$ has only a moderately smaller Fisher information --- about $0.36N$ \cite{BerWisBre01} --- and the minimum variance.

 In the asymptotic limit, 
\beq
{\rm HL} = V[\phi_{\rm est}]  \sim \ro{\pi/N}^2  \textrm{ for } N \gg 1,
\eeq
which is quadratically better than the SQL. The \hei-limit scaling implies that to obtain $K+1$ bits of precision for $\phi_{\rm est}$, we require of order $N=2^{K+1}$ qubits. That is, it is exponentially costly in ``spatial'' resources\footnote{In the context of optical interferometry, this corresponds to the exponential energy cost of NOON-like states discussed in Sec.~\ref{sec:IPE}.}. However this is not necessary. The fact that the optimal state and measurement can be written using a symmetrized basis with only $N+1$ basis states (out of a total Hilbert space dimension of $2^N$ for $N$ qubits) allows an alternate representation. Assuming $N=2^{K+1}-1$, we can use just $K+1$ qubits, and define a new $\ket{n}$ as the logical state of a register of qubits with $n$ a binary string of length $K+1$. The price to be paid is that the 
evolution generated by the phase gates, 
\beq
\sum_{n=0}^N \sin \left[ 
\frac{(n+1)\pi} {N+2}\right] \ket{n} \to  \sum_{n=0}^N e^{in\phi} \sin \left[ 
\frac{(n+1)\pi} {N+2}\right] \ket{n},
\eeq
now requires an exponential number of phase shifts on the ``most significant qubits'' in the binary representation of $n$. That is, we have swapped an exponential cost in spatial resources for an exponential cost in time resources, if each phase gate is assumed to take a fixed time.

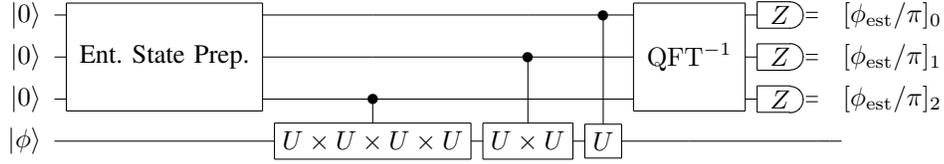
\begin{figure*}
$$
\Qcircuit @C=1.0ex @R=0.5ex @!R {
\lstick{\ket{0}} 	&\multigate{2}{\textrm{Ent.~State~Prep.}} &  \qw     			     & \qw     		& \ctrl{3} & \multigate{2}{{\rm QFT}^{-1}}  		&\measureD{Z}&\cw  
&\rstick{[\phi_{\rm est}/\pi]_0}\\
\lstick{\ket{0}}	&\ghost{\textrm{Ent.~State~Prep.}}& \qw     			     &\ctrl{2}	&\qw & \ghost{{\rm QFT}^{-1}}	  &\measureD{Z}& \cw  		 &\rstick{[\phi_{\rm est}/\pi]_1} \\
\lstick{\ket{0}}	&\ghost{\textrm{Ent.~State~Prep.}}& \ctrl{1}			     &\qw					&\qw 		        &\ghost{{\rm QFT}^{-1}}		 &\measureD{Z}& \cw  
&\rstick{[\phi_{\rm est}/\pi]_2}\\
\lstick{\ket{\phi}}        &	\qw		  &\gate{U\times U\times U\times U}  &\gate{U\times U}&\gate{U} & \qw  			     &\qw 		&\qw	 			 &\qw & \qw &\\
}
$$
\caption{Circuit representation of Heisenberg-limited interferometry using a binary encoding. Here 
the resource count is $N=4+2+1=7$.}
\label{fig:HL}
\end{figure*}
\begin{figure*}
$$
\Qcircuit @C=1.0ex @R=0.5ex @!R {
\lstick{\ket{0}} 	&\multigate{2}{\textrm{Ent.~State~Prep.}} &  \qw     			     & \qw    			& \gate{R(\smallfrac{\pi}{4})}		& \qw     				& \qw     				&  \gate{R(\smallfrac{\pi}{2})}    	& \ctrl{3}		&\measureD{X}&\cw  &&\\
\lstick{\ket{0}}	&\ghost{\textrm{Ent.~State~Prep.}}&  \qw     			     & \qw    			& \gate{R(\smallfrac{\pi}{2})}\cwx &\ctrl{2}			    &\measureD{X}&\control\cw\cwx 		 & \cw		& \cw  	&	\cw &&\rstick{\phi_{\rm est}} \\
\lstick{\ket{0}}	&\ghost{\textrm{Ent.~State~Prep.}}&\ctrl{1}			     &\measureD{X}&\control\cw\cwx 		 &\cw					&\cw						&\cw							&\cw				&\cw &\cw &\\
\lstick{\ket{\phi}}   & \qw     &\gate{U\times U\times U\times U}& \qw    	        & \qw  					     &\gate{U\times U}&\qw 				& \qw  						&\gate{U} &\qw &\qw  & \push{\phantom{2ex}} &\rstick{} 
\gategroup{1}{11}{3}{12}{0.2em}{\}}
\\
}
$$
\caption{Circuit representation of Heisenberg-limited interferometry as in Fig.~\ref{fig:HL}, but using the adaptive measurement scheme of Griffiths and Niu.}
\label{fig:GN}
\end{figure*}
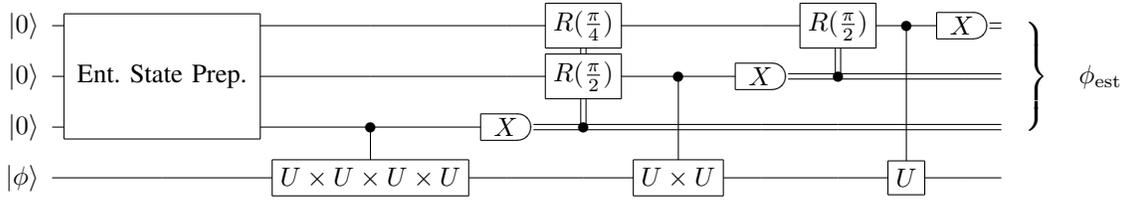
\begin{figure*}
$$
\Qcircuit @C=0.0ex @R=0.0ex @!R {
\lstick{\ket{+}}			&\push{\rule{1.5ex}{0.4pt}}\qw	&\qw					&\qw							&\qw     			&\qw							&\qw					&\qw				&\qw			&\qw						 &\gate{R(\theta)}			&\ctrl{3}	&\measureD{X}	&\control\cw 				& 		& \\
\lstick{\ket{+}}			&\qw							&\qw					&\qw							&\qw     			&\qw							&\gate{R(2\theta)}		&\ctrl{2}			&\measureD{X}	&\control\cw					 &\cwx					&			&				&\cwx 						&		& \\
\lstick{\ket{+}}			&\qw							&\gate{R(4\theta)}		&\ctrl{1}						&\measureD{X}	&\control\cw						&\cwx					&					&				&\cwx						 &\cwx					&			&				&\cwx 						&		& \\
\lstick{\ket{\phi}}			&\qw							&\qw\cwx				&\gate{U\times U\times U\times U}	&\qw 			&\qw\cwx						&\qw\cwx				&\gate{U\times U}	&\qw 			&\qw\cwx					 &\qw\cwx				&\gate{U}	&\qw			&\qw\cwx					&\push{\rule{0.8em}{0.4pt}\rule{0.8em}{0pt}}\qw 	&  \\
\lstick{\theta_{\rm init}}	&\cw							&\cw\cwx\blacklozenge	&\cw							&\cw      			& \cgate{D(\frac{\pi}{4})} \cwx	&\cw\cwx{\blacklozenge}	&\cw				&\cw     			&\cgate{D(\frac{\pi}{2})}\cwx	&\cw\cwx{\blacklozenge}   	&\cw		&\cw    		  	&\cgate{D(\delta\theta)}\cwx 	&\cw	&\rstick{\phi_{\rm est}}
}
$$
\caption{The Quantum Phase Estimation Algorithm for $K=3$, so that $N=7$.}
\label{fig:QPEA}
\end{figure*}
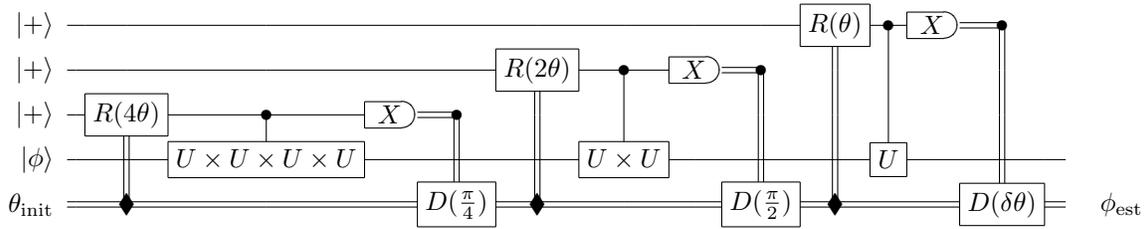

The canonical phase measurement, described by the phase states (\ref{phase_states}), 
is a measurement in a basis conjugate to the logical basis. The transformation from one 
basis to the other is exactly what is achieved by the quantum Fourier transform \cite{Sho94}. 
Thus using the binary representation rather than the symmetric representation, the \HL\ 
can be achieved by the circuit shown in Fig.~\ref{fig:HL}. In this instance, there are $K+1=3$ qubits, but 
$N=2^{K+1}-1 = 7$. The estimate $\phi_{\rm est}$ is read-out from the results of $Z$ measurements as shown, and we are using the notation $r=[r]_{0}.[r]_1[r]_2 \ldots$. The $k$th qubit ($k=0,1,\cdots, K$) ``passes'' the phase gate ${2^{k}}$ times. As stressed above, even though we represent (for instance) 4 applications of the phase gate by a single controlled-$U^4$ gate, this must be regarded as using 4 resources.

It is a remarkable fact, first pointed out by Griffiths and Niu \cite{Griffiths1996}, that the QFT$^{-1}$  can be achieved by local (i.e. single-qubit) {measurement and control}. This can be seen by moving the measurements back through the QFT$^{-1}$ and the controlled-phase gates, using the gate commutation properties. The control is often called feed-forward, but since each controlled 
qubit and measured qubit are entangled prior to the measurement, the control
is arguably feedback based on a partial measurement of a multi-qubit 
system. Indeed, we will use this terminology even when the qubits are independently prepared, 
because they are still {\em correlated} (from the point of view of the experimenter) due to the action of the 
phase gate with unknown phase $\phi$.   In any case, this adaptive scheme makes the measurement component of this \hei-limited protocol far easier to implement experimentally. See Fig.~\ref{fig:GN}.

\section{The Quantum Phase Estimation Algorithm}

Although the Griffiths-Niu technique makes the measurement easy to implement, attaining the exact 
 \hei\ limit still requires creating a multi-qubit entangled state, which is hard. This suggests exploring what happens if we replace the {entangled state} by independent qubits as in the standard scheme. This yields the {\em quantum phase estimation algorithm} (QPEA) \cite{Cleve1998,Nielsen00a}. If we mediate the control steps via the auxilliary phase $\theta$, and introduce a random $\theta_{\rm init}$ to ensure equal accuracy for all $\phi$, then the  QPEA is represented by the circuit in Fig.~\ref{fig:QPEA}.  This looks almost identical to the adaptive version of the standard protocol, as shown in Fig~\ref{fig:SQLii}. The difference is in the multiple gate applications on a single qubit; the QPEA with $K+1$ register qubits again uses ${N} = {2^{K}}+{2^{K-1}}+\cdots+{1} = 2^{K+1}-1$ resources, whereas the standard scheme uses $K+1$ resources.

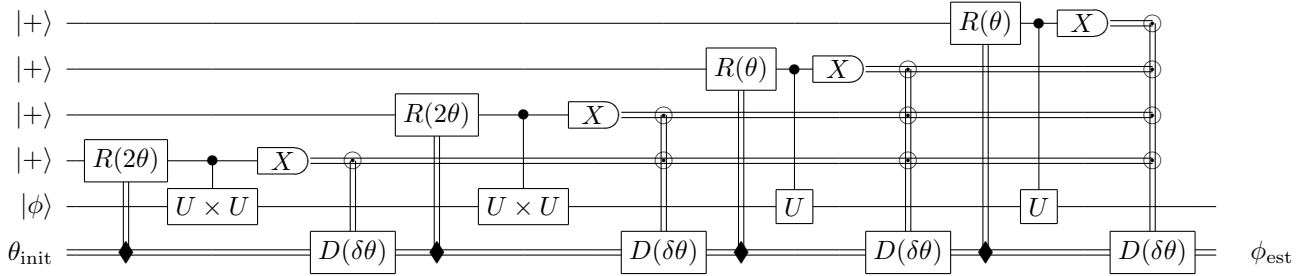
\begin{figure*}
$$
\Qcircuit @C=0ex @R=0.3ex @!R @W=12ex {
\lstick{\ket{+}}	&\push{\rule{1.5ex}{0.4pt}}\qw&\qw			&\qw		&\qw		&\qw					&\qw			&\qw		&\qw		&\qw					&\qw			&\qw		&\qw		&\qw					&\gate{R(\theta)}	&\ctrl{4}	&\measureD{X}	&\odot \cw &&&  \\
\lstick{\ket{+}}	&\qw&\qw			&\qw		&\qw		&\qw					&\qw			&\qw		&\qw		&\qw					&\gate{R(\theta)}	&\ctrl{3}	&\measureD{X}	&\odot\cw				&\cw\cwx		&\cw	 	&\cw		&\odot \cw \cwx &&& \\
\lstick{\ket{+}}	&\qw&\qw			&\qw		&\qw		&\qw					&\gate{R(2\theta)}&\ctrl{2}	&\measureD{X}	&\odot\cw				&\cw\cwx		&\cw		&\cw		&\cw\cwx\odot				&\cw\cwx		&\cw	  	&\cw		&\cw \cwx\odot &&& \\
\lstick{\ket{+}}	&\qw&\gate{R(2\theta)}	&\ctrl{1}	&\measureD{X}	&\odot\cw				&\cw\cwx		&\cw		&\cw		&\odot\cw\cwx				&\cw\cwx		&\cw		&\cw		&\cw\cwx\odot				&\cw\cwx		&\cw	 	&\cw		&\cw \cwx\odot &&& \\
\lstick{\ket{\phi}}	&\qw&\qw\cwx			&\gate{U\times U}	&\qw		&\qw\cwx				&\qw\cwx		&\gate{U\times U}  	&\qw		&\qw\cwx				&\qw\cwx		&\gate{U}	&\qw 		&\qw\cwx				&\qw\cwx		&\gate{U} 	&\qw		&\qw \cwx				&\push{\rule{0.8em}{0.4pt}\rule{0.8em}{0pt}}\qw& \\
\lstick{\theta_{\rm init}}	&\cw&\cw\cwx\blacklozenge		&\cw 		&\cw		&\cgate{D(\delta\theta)}\cwx	&\cw\cwx{\blacklozenge}	&\cw		&\cw		&\cgate{D(\delta\theta)}\cwx	&\cw\cwx{\blacklozenge}	&\cw		&\cw      	&\cgate{D(\delta\theta)}\cwx	&\cw\cwx{\blacklozenge}	&\cw		&\cw		&\cgate{D(\delta\theta)}\cwx	&\cw&\rstick{\phi_{\rm est}}\\ 
}
$$
\caption{Our generalized QPEA for the case ${M = 2}$ and $K=1$ (so that ${N = 6}$).}
\label{fig:GQPEA}
\end{figure*}

 Since the {QPEA} gives $K+1$ bits of $\phi_{\rm est}/\pi$, and $N\sim 2^{K+1}$ would we not expect 
\beq 
{\rm QPEA}:\ V[\phi_{\rm est}] \propto (\pi /2^{K+1})^{2} \sim (\pi/N)^{2}  \textrm{ (the HL) ?}
\eeq
Contrary to this expectation, an exact calculation \cite{Higgins07,Berry09} gives 
\beq {\rm QPEA}:\ V[\phi_{\rm est}] \sim 2/N  \propto 1/N \textrm{ (the SQL).}
\eeq
So what went wrong? Why does the algorithm not only fail to attain the \HL, but 
actually do worse than the SQL? The short answer is: Outliers. 
 The distribution $P(\phi_{\rm est})$ is sharply peaked around $\phi$. The half-width at half-maximum 
height is given by 
\beq
{\rm QPEA}:\ ({\rm HWHM})^{2} \simeq (2.81/N)^{2}.
\eeq
But the distribution has high wings, giving {SQL} scaling for the variance. Specifically \cite{Berry09}, 
\beq
P_{\rm QPEA}(\delta\phi) = \frac{\sin^{2}[N(\delta\phi/2)]}{N\sin^{2}[\delta\phi/2]},
\eeq
where $\delta\phi = \phi_{\rm est} - \phi$. In the wings, $P_{\rm QPEA}(\delta\phi)$ has an envelope that falls 
only like $(\delta\phi)^{-2}$. This is a consequence of the fact that we are using not the optimal state (\ref{psiWB}) with a wide but smooth number distribution, but rather a state with a ``flat'' number distribution,
\beq
\ket{\psi_{\rm flat}} = (N+1)^{-1/2} \sum_{n=0}^N \ket{n}.
\eeq
It is the sharp cut-off of the number coefficients that leads to the poor localization of phase, the conjugate variable.

Although outliers in $\phi_{\rm est}$ are important for phase estimation (where our figure of merit is the variance), they are not important for quantum computing applications. There, all one cares about is getting the right answer (to the number of bits of precision one has) 
from the algorithm with some reasonably high probability, and the QPEA works fine for this purpose \cite{Cleve1998,Nielsen00a}. Indeed, if $\phi/\pi$ had an exact binary expansion in $K+1$ bits, then we could remove the random $\theta_{\rm init}$, and the QPEA with $K+1$ qubits would be guaranteed to find $\phi$ exactly; the variance would be zero. While this assumption may be relevant in some quantum computing applications, it is contrary to the rules of the game (Sec.~III~A) for phase estimation.



\section{The Generalized QPEA} \label{sec:GQPEA}

One way to understand the high wings of the QPEA distribution is that if an error in an insignificant bit occurs, it propagates upwards into the more significant bits through the feedback protocol. This suggests a way to remove such errors: by repeating each measurement some number $M$ times, as previously suggested in various settings \cite{Kitaev1996,Rudolph2003,deBurgh2005,GLM06}. 
  Recall that in the {QPEA} the $k$th qubit ($k=0,1,\cdots, K$) controls the phase gate ${2^{k}}$ times. 
 We generalize this by having ${M}$ qubits for each ${2^{k}}$-fold application, so that the total number of passes through 
of the phase gate is
${N} = {M}\times{( 2^{K+1}-1)}$. 

With this generalization it is no longer clear how to change the auxiliary phase $\theta$ between measurements. Giovannetti, Lloyd, and Maccone \cite{GLM06}, considering the same problem as here, imply (when discussing a NOON-state realization rather than the equivalent binary-encoding implementation) that the adaptation of $\theta$  is unnecessary; we will return to this point in Sec.~\ref{sec:AAAN}. We prefer to keep the adaptation of $\theta$ because it is an integral part of the QPEA. We use the adaptive algorithm of Ref.~\cite{BerWis00} to make the {\em locally optimal} measurement, as explained in Sec.~\ref{sec:SQL}. For $M=1$  this {\em exactly} reproduces the {QPEA}, which is why we regard this family of algorithms (parametrized by $M$) as the natural generalization of the QPEA. 

For $M>1$ our generalized algorithm no longer realizes an optimal phase measurement; the Griffiths and Niu trick of realising an optimal phase measurement by local measurement and control only works for a single copy of the quantum register with binary-encoded phase. In fact, for $M=2$ as illustrated in Fig.~\ref{fig:GQPEA}, the measurement is far worse than optimal. As we have shown analytically \cite{Higgins07,Berry09}, the measurement in this case introduces so much noise that the estimate has a variance scaling at the SQL. (In this it is similar to the heterodyne measurement of Sec.~\ref{sec:optical_phase}.)  This is so despite the fact that the relevant state in this case, $\ket{\psi_{\rm flat}}^{\otimes 2}$, would give a nearly Heisenberg-limited phase estimate if one could implement an optimal measurement. For $M=3$, numerical simulations show that the non-optimal measurement introduces an excess noise variance with scaling consistent with $N^{-3/2}$, intermediate between the SQL and the \hei\ limit. (In this it is similar to the Mark II measurement of Sec.~\ref{sec:optical_phase}.) For $M\geq 4$, numerical simulations show that the measurement allows, as for the $M=1$ case\footnote{\hei-limited accuracy is not attained in the $M=1$ case because the prepared {\em state} does not have a \hei-limited phase variance.}, estimation with accuracy scaling at  the Heisenberg limit. All of these numerical simulations were performed up to $N > 10^6$, far into the asymptotic regime.

\begin{figure*}
$$
\Qcircuit @C=0ex @R=0.3ex @!R @W=12ex {
\lstick{\ket{+}}	&\push{\rule{1.5ex}{0.4pt}}\qw&\qw			&\qw		&\qw		&\qw			&\qw		&\qw		&\qw					&\qw			&\qw		&\qw		&\qw					&\gate{R(\theta)}	&\ctrl{4}	&\measureD{X}	&\odot \cw &&&  \\
\lstick{\ket{+}}	&\qw&\qw			&\qw		&\qw			&\qw			&\qw		&\qw		&\qw					&\gate{R(\theta)}	&\ctrl{3}	&\measureD{X}	&\cw				&\cw\cwx		&\cw	 	&\cw		&\odot \cw \cwx &&& \\
\lstick{\ket{+}}	&\qw&\qw			&\qw		&\qw		&\gate{ R(\theta)}&\ctrl{2}	&\measureD{X}	&\cw				&\cw\cwx		&\cw		&\cw		&\cw				&\cw\cwx		&\cw	  	&\cw		&\cw \cwx\odot &&& \\
\lstick{\ket{+}}	&\qw&\gate{R({2}\theta)}	&\ctrl{1}	&\measureD{X}	&\cw\cwx		&\cw		&\cw		&\cw				&\cw\cwx		&\cw		&\cw		&\cw			&\cw\cwx		&\cw	 	&\cw		&\cw \cwx\odot &&& \\
\lstick{\ket{\phi}}	&\qw&\qw\cwx			&\gate{U\times U}	&\qw	&\qw\cwx		&\gate{U}  	&\qw		&\qw			&\qw\cwx		&\gate{U}	&\qw 		&\qw			&\qw\cwx		&\gate{U} 	&\qw		&\qw \cwx				&\push{\rule{0.8em}{0.4pt}\rule{0.8em}{0pt}}\qw& \\
\lstick{\theta_{\rm init}}	&\cw&\cw\cwx\blacklozenge		&\cw 		&\cw			&\cw\cwx\blacklozenge	&\cw		&\cw		&\cgate{D(\smallfrac{\pi}{3})}	&\cw\cwx\blacklozenge	&\cw		&\cw      	&\cgate{D(\smallfrac{\pi}{3})}	&\cw\cwx\blacklozenge	&\cw		&\cw		&\cgate{D(\delta\theta)}\cwx	&\cw&\rstick{\phi_{\rm est}}\\ 
}
$$
\caption{Non-adaptive algorithm with $M(K,k) = {M_{K}} + {\mu}(K-k)$. Here $K=1$, ${M_{K}=1}$, and ${\mu = 2}$, so that $N=5$.}
\label{fig:NA}
\end{figure*}
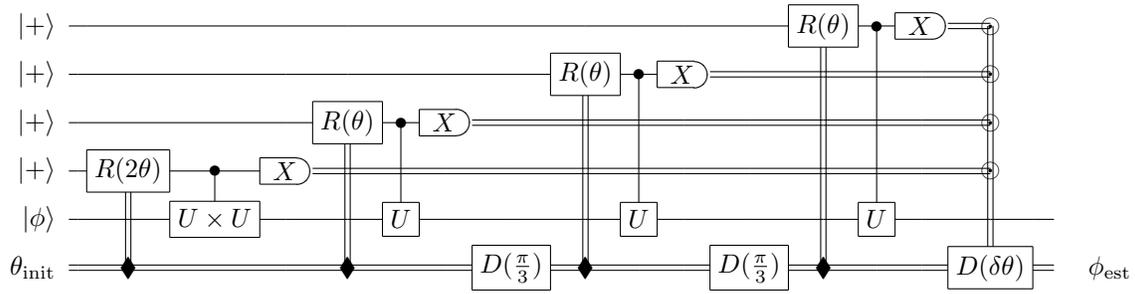

If one increased $M$ indefinitely, with $K$ fixed, one could not hope to achieve \hei-limited scaling. Intuitively, this is because the \hei-limited sensitivity comes from having a maximum  number of passes $p=2^K$ which scales linearly with $N$. Alternatively, it can be understood from the fact that  the state $\ket{\psi_{\rm flat}}^{\otimes M}$ ceases to have a broad number distribution (and hence a narrow phase distribution) as $M$ increases. In fact it can be shown analytically that for large $M$, the multiplicative overhead above the \hei\ limit increases like $M$. Thus there must be an optimal value of $M$ equal to 4 or higher. Numerically we found the best results to be for ${M = 5}$ \cite{Berry09}:
\beq
{M = 5}\ {\rm GQPEA}:\ V[\phi_{\rm est}] \simeq (4.8/N)^{2} , 
\eeq
compared to $(\pi/N)^{2}$ for the \HL. That is, even though the state $\ket{\psi_{\rm flat}}^{\otimes 5}$ is not the optimal state, and the locally optimal adaptive measurement is not a canonical phase measurement, the multiplicative overhead on $\Delta \phi$ is less than $1.53$. Although theoretically $M=5$ is the optimal choice for minimizing the variance, it still yields more outliers than is the case for larger $M$. Experimentally this means not that it is hard to measure with a precision scaling at the \hei\ limit, but rather that it is hard to prove this precision because of the difficulty of obtaining reliable statistics. For this reason, in Ref.~\cite{Higgins07} we experimentally implemented the generalized QPEA for $M=6$, for which the multiplicative overhead on $\Delta \phi$ is about $1.56$.

 \section{Are Adaptive Algorithms Necessary?} \label{sec:AAAN}

\subsection{A Less Adaptive ``Hybrid'' Algorithm}

 Our {Generalized QPEA} (Fig.~\ref{fig:GQPEA}) involves {\em Bayesian feedback}, in which every past result contributes to determining the auxilliary phase increment $\delta\theta$, according to the Berry--Wiseman protocol \cite{BerWis00}. This is to be contrasted with the original {QPEA} (Fig.~\ref{fig:QPEA}) in which only the {\em immediately preceding result} is required. As noted above, the QPEA would achieve the \HL\ if it were not for outliers. This suggests that we could  get to the HL more simply by augmenting the (simple adaptive) {QPEA} with the (non-adaptive) {SQL} algorithm to remove outliers. The reasoning is that the non-adaptive SQL algorithm, consisting basically of a large number of indepedent trials,  would be expected to have a probability distribution for the estimate which is roughly Gaussian, and so would have exponentially suppressed wings, and as long as a reasonable fraction of the resources is still devoted to the QPEA, then its narrow peak would hopefully remain intact.
 
 Investigating this hybrid scheme anaytically and numerically, reveals that  the optimal division of resources is $(2/3){ N}$ for the {QPEA} and $(1/3){ N}$ for the standard scheme \cite{Higgins09}. However, contrary to expectation, we find that 
\beq
{\rm Hybrid}:\ V[\phi_{\rm est}]  = w(N)/N^{3/2},
\eeq
where $w(N)$ is a function that increases very slowly with $N$ 
(from $4.83$ at $N=5$ to $6.17$ at $N=767$) \cite{Higgins09}. That is, the hybrid scheme delivers a scaling  {\em intermediate} between the  {SQL} and the \HL. Nevertheless it is interesting that by combining two measurement schemes, both of which give a variance scaling at the SQL, one obtains a much better scaling.

\subsection{A Non-Adaptive Local Algorithm}

The above hybrid algorithm is mostly non-adaptive --- it has only $K=O(\log N)$ adaptive measurements out of $O(N)$ measurements total. As noted, it surpasses the {SQL} in accuracy.  In it, the number of qubits with ${2^{k}}$ passes through the phase gate is ${M}(K,k) = 1 + \delta_{k,0}2^{K}$. This raises the question: can one get to the \HL\ with {\em no feedback} by choosing a {\em smoother} function ${M}(K,k)$ that still (like the hybrid scheme) assigns more qubits to smaller $k$-values?  Attacking this question analytically suggests considering functions of the form \cite{Higgins09} 
\beq
{M}(K,k) = {M_{K}} + \lfloor{\mu}(K-k)\rfloor, \label{MKk}
\eeq
where $\mu$ is a positive constant.  For each value of $k$, the $M(K,k)$ qubits are measured independently, using the auxilliary phase $\theta$ only to ensure an unbiased measurement. For example, it can be incremented by $\pi/M(K,k)$. This is illustrated in Fig.~\ref{fig:NA} with ${M_{K}=1}$ and ${\mu = 2}$.

 Numerically  we find the best results to be for ${M_{K}=2}$ and ${\mu=3}$ \cite{Higgins09}: 
\beq
\textrm{NALA}:\ V[\phi_{\rm est}] \simeq (6.4/N)^{2}.
\eeq
That is, the overhead on $\Delta \phi$  is less than $2.03$. This is not greatly bigger than 
the lowest known overhead for an adaptive scheme, of $1.53$ for the $M=5$ GQPEA. It is important to note that numerical simulations for a nonadaptive scheme with $M$ fixed do {\em not} show scaling at the \HL. This is contrary to the claim (rather casually made) in Ref.~\cite{GLM06} that the \HL\ for the variance can be attained from measuring each bit in the binary expansion of $\phi$ by making $\nu$ nonadaptive measurements (their $\nu$ is our $M$) for each bit, and then taking 
``the limit of large $\nu$''. It is not clear what this limit is meant to be;  $\nu$ certainly cannot be arbitrarily large as this would lead back to the SQL as discussed in Sec.~\ref{sec:GQPEA}.\footnote{Giovannetti {\em et al.} \cite{GLM06} perhaps miss this point because they appear to ignore the $\nu$ repetitions in their count of the resources, in deriving ``the Heisenberg limit $1/N$ [as] the ultimate bound to precision in phase measurements.'' Indeed they call $N$ the ``number of probes'' (i.e. register qubits), rather than the number of applications of the phase gate. This is despite the fact that they also say that ``Instead of a parallel strategy on $N$ probes, one can employ a sequential strategy on a single probe,'' while at the same time saying that   ``one finds the same  $1/N$ precision scaling \ldots for sequential strategies.''} Our numerics show that nonadaptive measurements with a large but fixed $M$ seem  to give a scaling close to the \HL\ up to a point, but as $N$ is increased (by increasing $K$) beyond that point the variance ceases to scale as $N^{-2}$.    

In additional to this numerical evidence, the above algorithm using $M(K,k)$ of the form (\ref{MKk}) is the only known nonadaptive algorithm on single qubits which has been rigorously proven to attain the \HL. 
The proof \cite{Higgins09,Berry09} involves values of ${M_{K}}$ and ${\mu}$ that are known (from numerics)  not to be optimal, but which allow one to rigorously bound 
all the contributions to the variance, using Chernoff's theorem. 
To apply this theorem one needs repeated {\em identical} measurements, 
so the proof assumes using just two values of $\theta$: $0$ and $\pi/2$. 
 Specifically, for ${M}(K,k) = 18 + \lfloor 16\ln(2)\times (K-k) \rfloor$ we prove that 
 \beq 
 \textrm{NALA}:\ V[\phi_{\rm est}] \lesssim (150/N)^{2}.
 \eeq
 The large overhead in this case shows that one would not want to use these parameters
 in practice, but it does prove rigorously that adaptive measurements are not necessary to 
 attain the \HL\ in interferometry using only single-qubit preparation and measurement.



\section{Conclusion}

Recent years have seen the addition  of adaptive measurements to the arsenal of techniques used in quantum optics information laboratories to probe the quantum world. The notable protocols which we have discussed here are: the Dolinar receiver \cite{Dol73}, realised in Ref.~\cite{JM07}; the Mark II phase measurement of Refs.~\cite{WisKil97,WisKil98}, realised in Ref.~\cite{Arm02}; and 
the generalized quantum phase estimation algorithm proposed and realized in Ref.~\cite{Higgins07}. 
These protocols are all intimately related to quantum information, through quantum communication, quantum computation, and quantum algorithm theory respectively. In this context, the recent quantum computing cluster-state experiment of Ref.~\cite{Jennewein07} should also be mentioned, as perhaps the first adaptive measurement in which the adapted measurement was performed on a subsystem that was initially entangled with the first-measured subsytem. 

The bulk of this paper has concentrated on the work of Ref.~\cite{Higgins07}, which is unique in that the 
adaptive measurement protocol was inspired by quantum algorithm theory, but serves a purpose quite different from quantum computing, namely estimating a completely unknown phase shift $\phi$ in one arm of an interferometer with a fixed number of photon-passes through the interferometer. We analysed this from a quantum information perspective (e.g. replacing photons by qubits and phase shifts by controlled-unitaries) to make the connection to quantum algorithm theory as explicit as possible. 

To attain exactly  the Heisenberg limit for the variance of the estimate, $V[\phi_{\rm est}] \sim (\pi/N)^{2}$, 
the most efficient (in terms of minimizing the number of qubits in the register, and the number of entangling gates performed on the register) protocol requires all of the following:
\begin{enumerate}
\item {preparing an entangled state} of $O(\log N)$ qubits.
\item {multiple applications of the controlled-unitary gate} by any given qubit.
\item adaptive measurements ({control} of individual qubits based on prior results).
\end{enumerate} 
We have shown numerically, and  experimentally in Ref.~\cite{Higgins07}, that using a generalized quantum phase estimation algorithm one can dispense with the entangled state preparation, and still achieve \hei-limited scaling, and indeed come very close to the \HL:
\beq
{\rm GQPEA}:\ V[\phi_{\rm est}] \sim (1.53\pi/N)^{2}.
\eeq
Given that it is impossible, using only single-qubit measurement and controls, to produce simultaneously the optimal state and the optimal measurement, our phase estimation algorithm must be close to the best achievable with this restriction. 

Finally, we briefly discussed some unpublished results \cite{Higgins09,Berry09} showing analytically and numerically that  one can achieve \hei-limited scaling, although with an increased overhead, even without the adaptive measurements. However this requires a sophisticated partitioning of resources, contrary to some claims in the literature  (e.g. Ref.~\cite{GLM06}).


%

\section*{Acknowledgment}

This work was supported by the Australian Research Council
(CE0348250, FF0458313, DP0986503). We acknowledge 
enlightening discussions with Andrew Doherty, who is a co-author with some of us 
on the paper in preparation mentioned in Sec.~\ref{sec:distinguish}, and with  
Morgan Mitchell, who is a co-author with us on the two papers 
in preparation  discussed in Sec.~\ref{sec:AAAN}.

\ifCLASSOPTIONcaptionsoff
  \newpage
\fi




\bibliographystyle{IEEEtran}
\bibliography{QMCrefsPLUS}%

%


\begin{IEEEbiography}[{\includegraphics[width=1in]{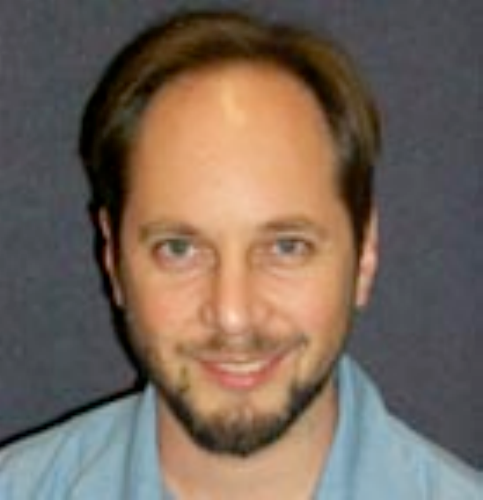}}]
{Howard Wiseman}  did his PhD in physics with Gerard Milburn at the University of Queensland (1994),  on quantum measurement and feedback. He has developed the control protocols used in a number of key experiments, including the first quantum-limited adaptive phase-estimation experiment. He is the author, with Gerard Milburn, of {\em Quantum Measurement and Control} (Cambridge, in production), the first comprehensive text book on this topic. 

Wiseman has over 130 journal papers and has received the Bragg Medal of the Australian Institute of Physics, the Pawsey Medal of the Australian Academy of Science, and the Malcolm Macintosh Medal of the Federal Science Ministry. He is a Fellow of the Australian Academy of Science and Director of the Centre for Quantum Dynamics of Griffith University, Brisbane.\end{IEEEbiography}







\end{document}